\documentclass[12pt]{article}

\usepackage{color}
\usepackage{fullpage}
\usepackage{natbib}
\usepackage{pgfplots}
\usepackage{tikz}
\usepackage{url}

\usepackage{xcolor}
\definecolor{tonyblue}{RGB}{0, 102, 204}

\begin{document}

\title{Grand Challenges in Bayesian Computation}
\author{Anirban Bhattacharya$^1$, Antonio Linero$^2$, Chris. J. Oates$^{3,4}$ \\
\small $^1$ Texas A\&M University, US  \\
\small $^2$ University of Texas as Austin, US  \\
\small $^3$ Newcastle University, UK  \\
\small $^4$ Alan Turing Institute, UK  
}
\date{}
\maketitle

\begin{abstract}
This article appeared in the September 2024 issue (Vol. 31, No. 3) of the Bulletin of the International Society for Bayesian Analysis (ISBA). 
\end{abstract}

Computation is arguably one of the fastest evolving subfields of Bayesian statistics at the moment, driven by a combination of democratised access to computing technologies (such as automatic differentiation) and recent algorithmic advancement. 
While the many successes of Bayesian computation are well-publicised at conferences and in journals, the open questions and problems of pressing importance are not so frequently discussed. 

Almost a decade ago, \citet{green2015bayesian} summarised the state-of-the-art in Bayesian computation, focusing primarily on algorithmic advances in Markov chain Monte Carlo, approximate Bayesian computation, and proximal gradient methods.
To shed some light on the current situation, we polled the current membership of the Computation section of ISBA. Here we present both a summary of these results, together with our own view of the current ``grand challenges" for Bayesian computation in 2024.  

As an opening gambit, participants were asked ``How significant is the challenge of computation in the context of Bayesian statistics?''.  All responders agreed that computation is a significant challenge, while it was interesting that 60\% of responders viewed computation as not the most significant challenge in Bayesian statistics at the moment.  This may reflect the success of the community to-date in developing computational solutions to facilitate Bayesian analyses, but may also reflect the other well-known challenges in the Bayesian workflow \citep{gelman2020bayesian,gelman2020holes}. 

\begin{center}
\pgfplotsset{width=7cm,compat=1.18}
\begin{tikzpicture}
\begin{axis}[
    ybar,
    title style = {align = center},
    title={How significant is the challenge of computation \\ in the context of Bayesian statistics?},
    enlargelimits=0.25,
    ylabel={\#responses},
    symbolic x coords={most significant,significant,not significant},
    xtick=data,
    ymin = 3,
    nodes near coords,
    nodes near coords align={vertical},
    x tick label style={rotate=25},
]
    \addplot coordinates {
        (most significant,10) (significant,15) (not significant,0)
    };
\end{axis}
\end{tikzpicture}
\end{center}

Next, participants were asked ``What do you see as being the most important class of computational methods for facilitating Bayesian statistics in 5 years' time?''.  An overwhelming 44\% of responders identified sampling methods (i.e. based on Monte Carlo) as most important, with amortised posterior approximation methods second on 24\%, and nonparametric variational approximations third on 12\%.  The preference for sampling methods may in part be due to their current widespread usage and existing software support, while it was interesting to see the nascent areas of amortised and nonparametric variational approximation enjoying perceived potential. 
Outside of these methods, one respondent noted some overlap between these different strategies, and one pointed to exact sampling methods as having potential.

Given the rapid advances in Machine Learning and Artificial Intelligence (AI), we asked how these are ``likely to interplay with Bayesian statistics in the next 5 years''.   While responses were diverse, a common theme was the application of flexible distribution approximation methods, such as diffusion models and normalising flows, to the posterior approximation task.  Another common theme was the use of AI in the Bayesian workflow, from using chatbots for prior elicitation from experts, to the use of AI assistants for conducting the statistical analysis itself.  Some responders also partly objected to the question, arguing instead for an increased role of Bayesian statistics in Machine Learning and AI! Overall, roughly two thirds of respondents expressed positive sentiments about the role of AI in Bayesian computation, with almost all other respondents expressing uncertainty. Exemplar positive and unsure comments are (positive):
\begin{quote}
  \it
  Without a doubt machine learning algorithms will play an important role in advancing the field of Bayesian computation. This is already happening with neural networks being used in MCMC algorithms and posterior inference with denoising diffusion models as just two examples. There were quite a few talks at the ISBA conference this year which also illustrated these connections between machine learning and Bayesian statistics. Going forward for the next 5 years, I think we will see Bayesian statisticians increasingly using machine learning algorithms. But hopefully, Statisticians will be able to contribute some novelty to this intersecting field and not just become users of these techniques. For example, there's a lot of missing theoretical understanding in machine learning and this is an opportunity for Statisticians to play a role in filling-in that theoretical gap.  
\end{quote}
and (unsure):
\begin{quote}
  \it
  The interplay will be significant in certain areas of application and much smaller in others. It is paramount that the community doesn't place all eggs in one basket and continues to work on new directions, more divorced from hype. 
\end{quote}

The final question we asked was ``What algorithms or features would you like to see being incorporated into new or existing off-the-shelf software for Bayesian computation?''. This free-form question delivered the most diverse response set, from which we pick three examples to highlight: software support for the full Bayesian workflow, ability to automatically differentiate through marginal likelihood, and improved software support for sequential Monte Carlo.

This brings us to the issue of \emph{grand challenges} for Bayesian computation. 
Section members were asked to identify grand challenges in free text, and responses from the community were predictably diverse, covering scalability of methods to large models and datasets, better leveraging of automatic differentiation and GPU acceleration, better sampling of complex/multimodal distributions, accurate computation of model evidence, among many others topics.
Taking inspiration from these responses, we have identified three specific grand challenges for Bayesian computation, applicable to the field as it stands in 2024.
These challenges below are the views of the authors and should not be interpreted as the views of the section as a whole.

\paragraph{Grand Challenge 1:  Understanding the Role of Parametrisation}

At the start of the decade, the strong generalisation performance often observed in deep learning was mainly attributed to the implicit regularisation afforded by stochastic gradient descent, and considerable research effort was devoted to improving stochastic optimisation algorithms and understanding their inductive biases.  However, a relatively recent paradigm shift has occurred, with favourable inductive biases now mainly attributed to how the neural architecture is parametrised.  That is, the loss landscape induced by a particular parametrisation strongly determines the local minima found by \emph{any} stochastic optimisation method, and the generalisation performance of the associated solution, motivating research into understanding the implications of how a neural network is parametrised.  It is interesting to observe that a wide range of tasks in Machine Learning are now tackled with variations of the same transformer architecture \citep{goldblumposition}, supporting this viewpoint.  Our conjecture is that a similar paradigm shift is needed in Bayesian computation; the community has been prioritising the development of new algorithms over understanding when existing algorithms work well, and how their performance can be improved through more careful consideration of how the statistical model is parametrised.  Indeed, many of our survey participants cited the development of improved algorithms as a grand challenge, and it is common to read research reporting that an algorithm was found to perform well or poorly on a particular posterior approximation task without consideration that performance can depend on how the posterior is parametrised.  A promising line of research could be to identify pairings of model parametrisations and algorithms for which the posterior approximation works well; the accumulation of these data would enable identification of broad guidelines or principles to inform how a model should be parametrised. 
Examples of work in this direction include \cite{van2001art} on the impact of data augmentation strategies on the EM algorithm/Gibbs sampler, and  \cite{papaspiliopoulos2007general,yu2011center,betancourt2015hamiltonian} on the role of centred versus non-centred parameterisations in hierarchical models for the purpose of boosting the performance of MCMC. 

\paragraph{Grand Challenge 2:  Community Benchmarks}

Bayesian computation has historically lacked a systematic approach to comparing different algorithms, with test problems often being cherry-picked to demonstrate the effectiveness of a proposed method \citep{chopin2017leave}.  Our view is that this practice betrays the evidence-based reasoning that we would as statisticians seek to promote in an applied context, hinders the identification of promising research directions, and falls short in respect of scientific rigour in comparison to related fields such as Machine Learning.  
Indeed, \citet{green2015bayesian} anticipated ``a threat that the whole field turns into a library of machine-learning techniques, with limited validation on
reference learning sets and a quick turnover of methods, which would both impoverish the field and fail to reach a general audience of practitioners''.
Though we do not share the same attitude toward Machine Learning, we are equally supportive of recent attempts to develop community benchmarks, such as the \texttt{posteriordb} benchmark developed by \citet{magnusson2024posteriordb}.  
The availability of a common set of test problems, together with a gold-standard ground truth, is an essential prerequisite to comparing the performance of the litany of different algorithms that are now available.  However, there is still much work to be done in this respect.  Notably, identifying test problems for which a high quality ground truth is available is difficult (e.g. \texttt{posteriordb} relies on an extended run of the No U-Turn Sampler), and instabilities in automatic differentiation currently preclude the plug-and-play use of such benchmarks without additional engineering work.  
The broader adoption and critical discussion of benchmark test problems by the community \citep[e.g.][]{heaton2019case} would surely catalyse further development of valuable community benchmarks.

\paragraph{Grand Challenge 3: Reliable Assessment of Posterior Approximations}

A recurring theme in survey responses was the need for better tools --- both theoretical and practical --- for assessing whether or not a particular approximation of the posterior distribution is fit for use. This includes creating diagnostic tools both for quickly and accurately measuring the quality of approximations of posterior distributions \citep{vehtari2021rank, yao2024discriminative} but also establishing theoretical guarantees on (for example) variational approximations when used for specific purposes (\citealp{wang2019frequentist, yang2020alpha} for parameter estimation; \citealp{zhang2024bayesian, ray2022variational} for model selection consistency). On the theoretical side, important subproblems include (i) providing tight, computable, bounds on approximation error of approximate posteriors (as in \citealp{huggins2020validated}) with either finite sample or asymptotic guarantees, and (ii) establishing that approximate posteriors, while possibly deficient as approximations to the true posterior, may nevertheless possess properties that make them reliable for specific problems such as model selection or uncertainty quantification for low-dimensional functionals of interest. 
We believe part of the reason sampling methods enjoy popularity over alternatives is because the guarantees they possess are better understood, more trusted, and (asymptotically) stronger than those that exist for non-sampling methods; narrowing this gap, either in terms of the approximation error for the full posterior or in terms of specific quantities like marginal likelihood approximations or other marginals of interest, would therefore make it easier to sell non-sampling methods to users who are interested in reliable uncertainty quantification.

\bigskip

It is of course not possible to summarise the challenges of Bayesian computation in terms of a small number of well-posed problems, as the wide range of responses to our survey testified.
Nevertheless, we feel it is valuable to highlight these three particular challenges for discussion, in the hope that new ideas and techniques can be developed that in turn will help to advance our field.

\bigskip

\noindent The authors wish to thank all of the members of the Computational section of ISBA who voluntarily took part in this survey.

\end{document}